\documentclass{article}
\usepackage{spconf,amsmath,graphicx,epsfig,epstopdf}
\usepackage{verbatim}
\usepackage{bm}

\title{On Investigation of Unsupervised Speech Factorization Based on Normalization Flow}
\name{Haoran Sun, Yunqi Cai, Lantian Li, Dong Wang}
\address{CSLT, Tsinghua University, China}

\begin{document}
%\ninept
%
\maketitle
\begin{abstract}

Speech signals are complex composites of various information, including phonetic content, speaker traits,
channel effect, etc. Decomposing this complicated mixture into independent factors, i.e., speech factorization,
is fundamentally important and
plays the central role in many important algorithms of modern speech processing tasks. In this paper, we present a preliminary
investigation on unsupervised speech factorization based on the normalization flow model. This model constructs a complex invertible transform,
by which we can project speech segments into a latent code space where the distribution is a simple diagonal Gaussian.
Our preliminary investigation on the TIMIT database shows that this code space exhibits favorable properties such as
denseness and pseudo linearity, and perceptually important factors such as phonetic content and speaker trait
can be represented as particular directions within the code space.

\end{abstract}
\begin{keywords}
speech factorization, normalization flow, deep learning
\end{keywords}
\section{Introduction}
\label{sec:intro}

Speech signals are complex composite that involves enormous information, such as
phonetic content, speaker traits, emotion, channel and ambient noise. These different types of
information are intermingled in an unknown manner, which leads to the fundamental difficulty
in all speech processing tasks.
For example, speaker variation is among the most challenging problems
that speech recognition researchers have been annoyed for several decades, and the variation on
emotion status and speaking styles causes notorious trouble for speaker recognition~\cite{benesty2007springer}.

A natural idea to deal with the information blending is to factorize the speech signal into
separate informative factors. This idea has been partly demonstrated by some well-known
factorization-based models in speaker recognition, such as JFA~\cite{Kenny07} and the
i-vector model~\cite{dehak2011front}, where speech signals are assumed to be factorized
into phonetic content, speaker traits and channel effect. A clear shortage of these
methods is that they assume the informative factors are composed by a linear Gaussian model,
which might be oversimple and is incapable of describing the complex generation process of speech signals.

Recently, we proposed a deep cascade factorization (DCF) approach to factorize speech signals at
the frame level~\cite{li2018factor}. The DCF approach follows the layer-wised generation process
proposed by Fujisaki~\cite{fujisaki1997prosody} and factorize speech signals into
information factors one by one, and each new factor depends on the factors that have been
inferred already, by using a task-oriented deep neural network (DNN) trained using task-specific data.
DCF is the first successful speech factorization model based on deep learning, however it
suffers from two shortcomings: (1) it is based on supervised learning and requires labelled data
for all interesting information factors; (2) it is frame-based and does not consider the temporal
dependency within speech signals.

In this paper, we present an unsupervised speech factorization approach based on deep generative
models. The basic hypothesis is that if we can find a way to generate the data, then we probably
can gain a better understanding of the underlying informative factors. The i-vector model is such a generative model, but
the linear Gaussian assumption is too strong to suit the generation process of speech. In this study,
we make use of the powerful generation capability of DNNs to deal with this problem.
More specifically, we build a latent code space where the distribution is as simple as a diagonal Gaussian,
and train a complex DNN to generate the speech signals from these latent codes. We found that
perceptually important speech factors could be represented as particular directions within the code space.

%By this
%approach, the underlying factors of speech signals and the generation process can be separated,
%and we analyze and manipulate speech signals in the factor space.

There are three popular deep generative models: generative adversarial network (GAN)~\cite{goodfellow2014generative}, variational
auto-encoder (VAE)~\cite{kingma2013auto} and normalization flow~\cite{dinh2014nice,dinh2016density,rezende15variational,Kingma2016Improving}.
Among these models, GAN is capable
of generation but lack of inference. VAE is capable of both generation and inference, but
the model is trained with a variational bound of the true likelihood, hence not accurate.
Normalization flow is trained to maximize the true likelihood, and the generation and inference are
simple. This model has been successfully applied to image generation~\cite{kingma18glow}
and speech synthesis~\cite{oord18parallel,Kim2018FloWaveNet,Prenger2019Waveglow}.
In this work, we extensively use the normalization flow model to study the speech generation process
and investigate the property of the latent code space.

\section{Related work}

The idea of discovering and manipulating speech factors plays the central role in
many important algorithms in speech processing. The most important example is the famous source-filter model
and the associated linear prediction coding (LPC) algorithm, which decomposes speech signals into vocal fold excitation and
vocal tract modulation~\cite{Fant1960acoustic,Atal2006The}. This decomposition places the foundation of modern
speech processing theory, however, it is mostly psychologically inspired and the factors it
derives (excitation and modulation) are not directly related to speech processing tasks.
For example, neither excitation nor modulation directly represents speaker traits.
By contrast, the Fujisaki model~\cite{fujisaki1997prosody} treats speech
generation as a process of convolution of different layers of informative factors, and each factor is
related to a specific speech processing task. However, the inference with Fujisaki model is difficult.
The DCF algorithm~\cite{li2018factor}
provides the inference approach, however, the model training requires a large amount of labelled data. This paper
presents an unsupervised approach that can train a factorization model with unlabelled data
and infer informative factors in an easy way.

This work is related to the flow-based speech synthesis~\cite{Kim2018FloWaveNet,Prenger2019Waveglow},
but our goal is to analyze rather than generate speech signals. The unsupervised factorization idea
was also seen in recent work on multi-speaker and multi-style speech synthesis~\cite{wang2018style,hsu2019}.
Finally, this work is mostly inspired by the flow-based image generation~\cite{kingma18glow}.

%Fortunately, our recent study showed that speaker traits are largely short-time spectral
%patterns, and a carefully designed deep neural network can learn to extract these patterns
%at the frame level~\cite{li2017deep}.
%The following studies demonstrated that the frame-level deep speaker features are highly generalizable: they
%work well with voices of trivial events, such as laugh and cough that are as short as 0.3 seconds~\cite{zhang2017speaker}; and they
%are robust against language mismatch~\cite{li2017cross}. The short-time property of speaker traits suggests
%that speech signals are possibly short-time factorizable, as it has been known that another major speech factor,
%the linguistic content, is also short-time identifiable~\cite{hinton2012deep}.

\section{Normalization flow for speech factorization}
\label{sec:model}

\subsection{Review on normalization flow}

\begin{figure}[htbp]
\centering\includegraphics[width=0.90\linewidth]{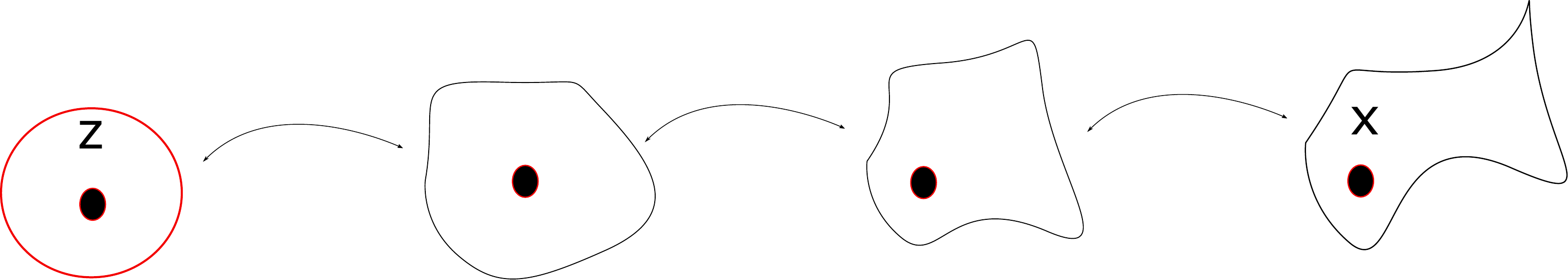}
\caption{Normalization flow transforms a simple distribution of $z$ to a complex distribution of $x$, by a series of
invertible transforming functions.}
\label{fig:flow}
\end{figure}

The basic idea of normalization flow is to design a chain of invertible transforms that map a simple distribution to
a complex distribution~\cite{dinh2014nice}, as shown in Fig.~\ref{fig:flow}.
In this figure,  each single-step transform is invertible, so the whole transform is invertible.
By this invertible transforming, a variable $z$ that follows a simple distribution can be mapped to a variable $x$
whose distribution is very complex. Conversely, a variable $x$ whose distribution is complex can be mapped back to
a variable $z$ whose distribution is simple. This transform chain is called a \emph{normalization flow}.
In our study, we will treat $x$ as a speech signal, and our goal is to transform the signal to a code $z$, which
encodes the latent factor underlying $x$.

According to the principle of distribution transforming~\cite{watkins2009}, the probabilities
of the observation $x$ and the corresponding code $z$ possess the following relation:

\[
\ln p(x) = \ln p(z(x)) + \ln |\text{det}(\frac{{\rm d} z} {{\rm d} x})|,
\]

\noindent where $z(x)$ is the inverse function of $x(z)$, and $\text{det}(\frac{{\rm d} z} {{\rm d} x})$ is the
determinant of the Jacobian matrix $\frac{{\rm d} z} {{\rm d} x}$.

The flow can be trained following the maximum likelihood (ML) criterion, for which the objective function can be written as follows:

\[
L_{\theta} = \sum_i \ln p(x_i) = \sum_i \ln p(z_{\theta}(x_i)) + \ln \frac{{\rm d} z_\theta(x_i)} {{\rm d} x_i}.
\]

\noindent where $\theta$ denotes the parameters of the flow. Maximizing this objective leads to a deep
generative model where the generation process simply casts to sampling $z$ from $p(z)$ and transforming it to the observation space by $x(z)$.
Conversely, the (inverse) flow can be used to transform an observation $x$ to its code $z$, offering a tool to describe data in the
code space.

Note that normalization flow is a general framework, where both the transforming functions and the latent distribution can be selected freely.
Regarding the transforming function, popular choices are linear transform~\cite{rezende15variational}, inverse autoregressive transform~\cite{Kingma2016Improving},
and $1 \times 1$ convolutional transform~\cite{kingma18glow}. All these transformations are invertible and the associated Jacobians hold simple forms.
Regarding the latent distribution, the most popular choice is the diagonal Gaussian.

\subsection{Speech factorization by flow}

With the normalization flow model, it is possible to transform a speech signal $x$ to a latent code $z$ whose distribution is
as simple as a diagonal Gaussian. Since the distribution of $z$ is much simpler than $x$, it becomes easier for us
to analyze speech signals in the code space, paving the way of discovering important informative factors there.

A key concern of speech factorization is the dependency over time. Although frame-based factorization worked in our
previous work~\cite{li2018factor}, we suppose taking temporal dependency into account would help. Therefore,
the data sample $x$ we choose in this study is a speech segment rather than a speech frame, the length of which
is fixed. We first compute these spectrograms of the speech segments, and then treat these spectrograms as
observations in the flow model. Considering that the spectrograms are 2-dimensional images, we choose the Glow structure~\cite{kingma18glow}
to implement the normalization flow, as it has worked well in image generation tasks.

As a preliminary study, we choose to analyze English vowels. The goal of the analysis is to study the distributional properties of
different vowels and speakers in the code space, and investigate the possibility to discover important factors
that are perceptually salient for human ears.

\section{Experiments}

\subsection{Data preparation}

We use the TIMIT database to conduct the experiments, and choose five English vowels (aa, ae, iy, ow, uh) in the investigation. Firstly, the speech segments of the five target vowels are extracted from the TIMIT database according to the meta information of the speech utterances.
Secondly, these speech segments are converted to spectrograms, by setting the window length, window hop and FFT length to be $25$ms, $1$ms, $257$, respectively.
Thirdly, spectrograms longer than $288$ frames are discarded,
and all the rest spectrograms are lengthened to $288$ frames by appending zeros.
This leads to spectrograms in size of $288 \times 288$ pixels, which are used as the observations of the Glow model.
The first $288$ denotes the number of frames in the time domain,
and the second $288$ denotes the number of frequency bands in the frequency domain with a dimension of FFT length appending 31 zeros.
%These fixed-length speech segments
%are then converted to spectrograms, by setting the window length, window hop and FFT length to be 25ms, 1ms, 257, respectively.

We use the code from an available version of the Glow model to conduct training and inference\footnote{https://github.com/chaiyujin/glow-pytorch}.
All the audio examples reported in the following experiments can be downloaded from http://project.cslt.org.

\subsection{Distribution of observations and codes}

\begin{figure}[htbp]
\centering\includegraphics[width=0.82\linewidth]{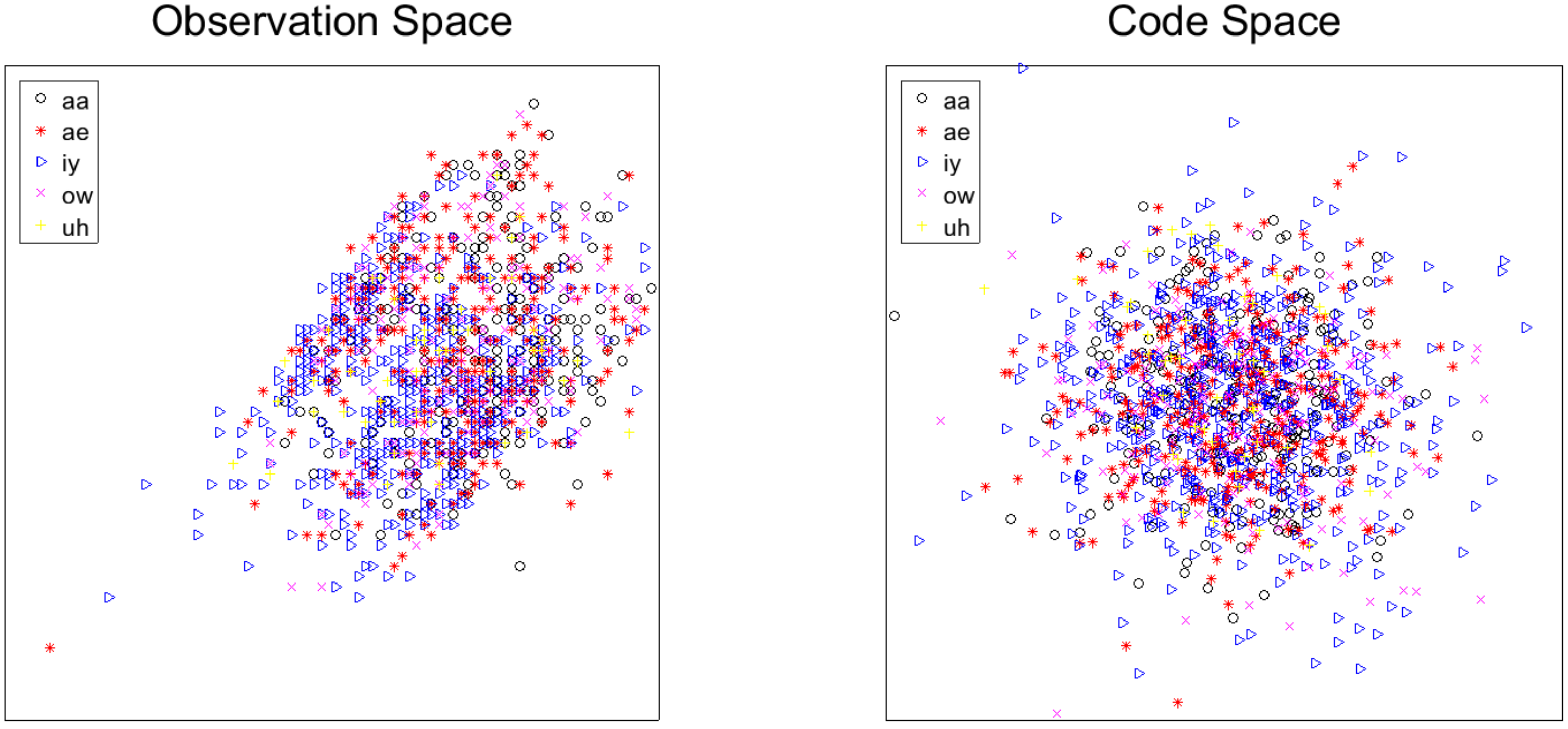}
\caption{Distribution in the observation space (left) and the code space (right).}
\label{fig:dist}
\end{figure}

Fig.~\ref{fig:dist} shows the distribution of the observations and codes, where we randomly select two dimensions for each piece of data.
It can be seen clearly that the distribution of the codes is much more Gaussian compared to the distribution of the
observations. This verified that the flow model has been well trained and it indeed normalized the distribution.

\vspace{-1mm}

\subsection{Sampling}

\begin{figure}[htbp]
\centering\includegraphics[width=0.94\linewidth]{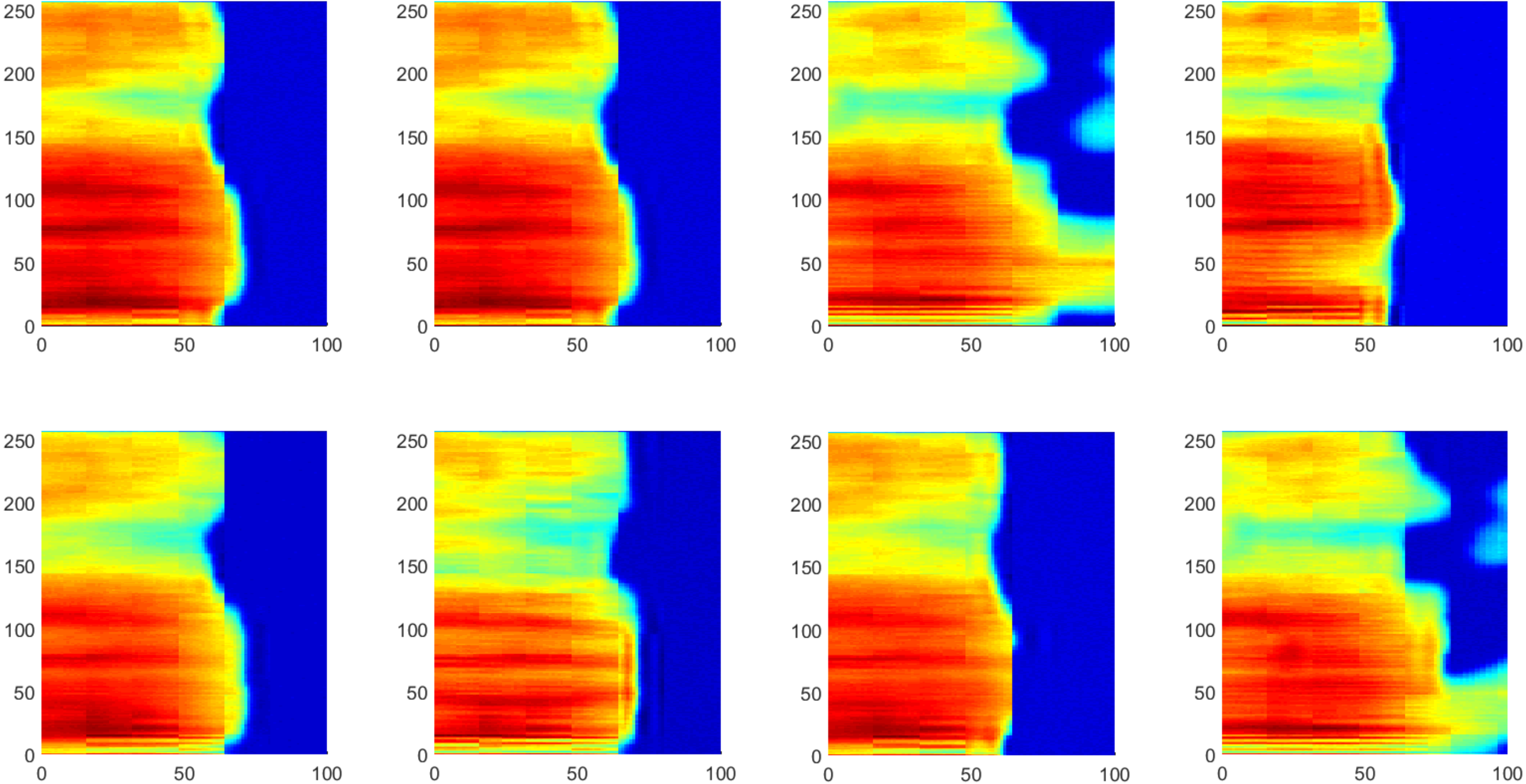}
\caption{Spectrograms obtained by sampling in the code space and transforming to the observation space.}
\label{fig:sample}
\end{figure}

In this experiment, we test the flow by sampling some speech segments in the code space.
This can be achieved by sampling a code $z$ following a diagonal Gaussian, and then
transforming it to an observation $x$ (spectrogram) through the flow $x(z)$. Fig.~\ref{fig:sample}
shows some spectrograms of sampled examples. It can be seen that the sampled spectrograms exhibit
similar formant structures to those of true speech. By converting them to waveforms using phase of a
true speech, we found these samples are meaningful speech.
This is not surprising as the distribution of
meaningful speech segments is Gaussian in the code space, so samples obtained
following the Gaussian will have a high probability to be meaningful.
An observation is that most of the samples we obtained are silence.
This could be attributed to the large proportion of silence in the training data,
caused by the silence padding.

\vspace{-2mm}

\subsection{Interpolation}

In this experiment, we investigate the (pseudo) linear property of the code space.
Considering two meaningful speech segments, both
should be located in a \emph{dense} area as they are in the code space.
This is because they are meaningful and so should be granted high probabilities by the model.
According to the property of the diagonal Gaussian, in the code space,
the probability at the location of any interpolation of the two segments
will be between the dense locations of these two segments. This means that any interpolation
will result in a meaningful speech segment. Ideally, the speech properties will change gradually from
one segment to the other by the interpolation.

\begin{figure}[htbp]
\centering\includegraphics[width=0.94\linewidth]{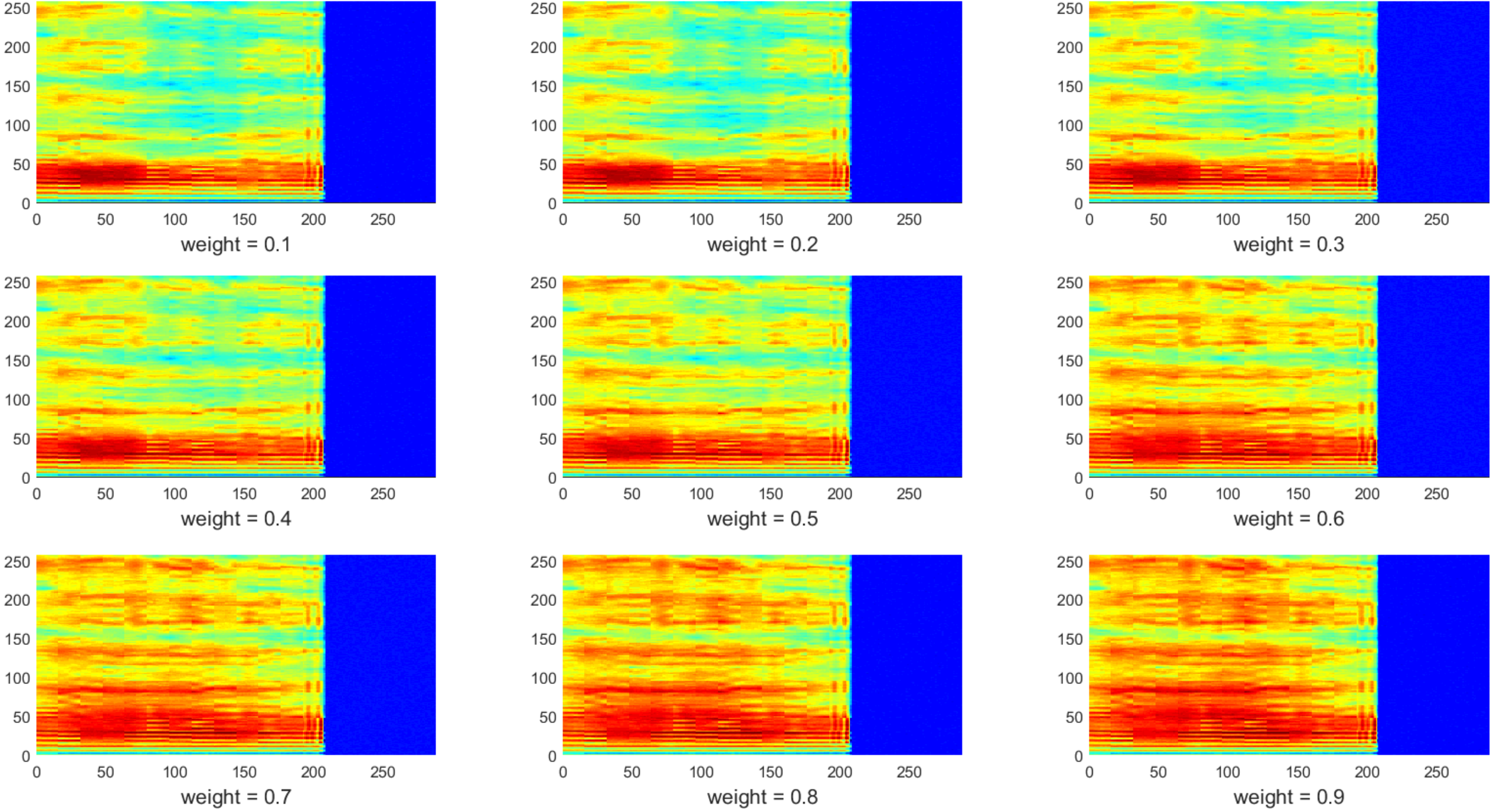}
\caption{Interpolation between vowel $aa$ and $ae$ in the code space. From top-left
corner the bottom-right corner, the interpolation weight for $ae$ changes from $0.1$ to $0.9$.}
\label{fig:interp}
\end{figure}

To test this hypothesis, we choose a segment of $aa$ and a segment of $ae$ spoken by the
same person, and interpolate them in the code space. Results are shown in Fig.~\ref{fig:interp}.
It is interesting that by this interpolation, a segment of $aa$ gradually
changed to a segment of $ae$, without much change on other properties, e.g., speaker traits.
The audio clips reconstructed from the spectrograms (by using the phase of $aa$) can be
downloaded online; they sound rather reasonable. We also test interpolation between
genders and speakers, both work well.

This result is highly interesting, as it suggests that the code space is likely
pseudo linear for factors that are salient for human ears. In other words, in the code
space, it is possible to find a direction following which only one perceptually important
factor changes. An implication of this property is that a speech factor can be represented by a
particular direction in the code space.
This suggests a possible speech factorization strategy that starts from a neutral speech,
and change its properties by moving in the code space following the directions that correspond to
the desired properties sequentially.

\vspace{-1mm}
\vspace{-1mm}

\subsection{Denoising}

\begin{figure}[htbp]
\centering\includegraphics[width=0.94\linewidth]{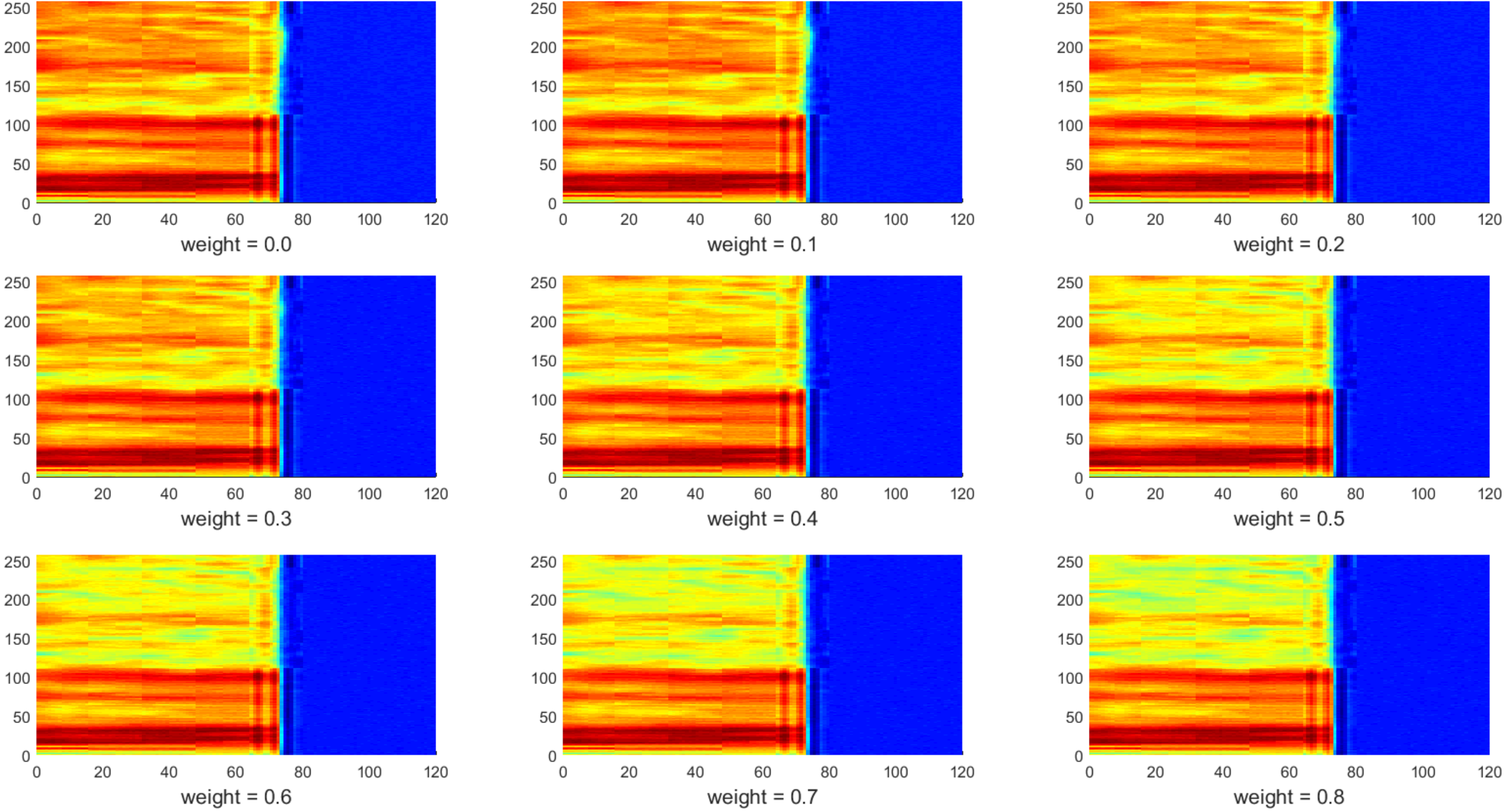}
\caption{Denoising using the pseudo linear property of the code space. From the top-left corner to bottom-right corner, the denoising scale $\beta$
changes from 0 to 0.8.}
\label{fig:denoise}
\end{figure}

The pseudo linear property of the code space can be used to remove noise.
Firstly we add white noise to the training data
randomly, and then compute the codes for these noise-contaminated segments. The averaged
codes for clean and noisy speech are computed respectively, and the displacement
between them, denoted by $\xi_{c \rightarrow n}$, is used to recover the clean version for noise-corrupted segments,
by $z_c = z_n - \beta \xi_{c \rightarrow n}$, where $\beta$ is a denoising scale.
The effect of the factorization-based denoising is shown in Fig.~\ref{fig:denoise}. It can be
seen that noise is removed gradually when stepping towards the opposite direction of $\xi_{c \rightarrow n}$.

\vspace{-1mm}
\vspace{-1mm}

\subsection{Discrimination}

\begin{figure}[htbp]
\centering\includegraphics[width=0.92\linewidth]{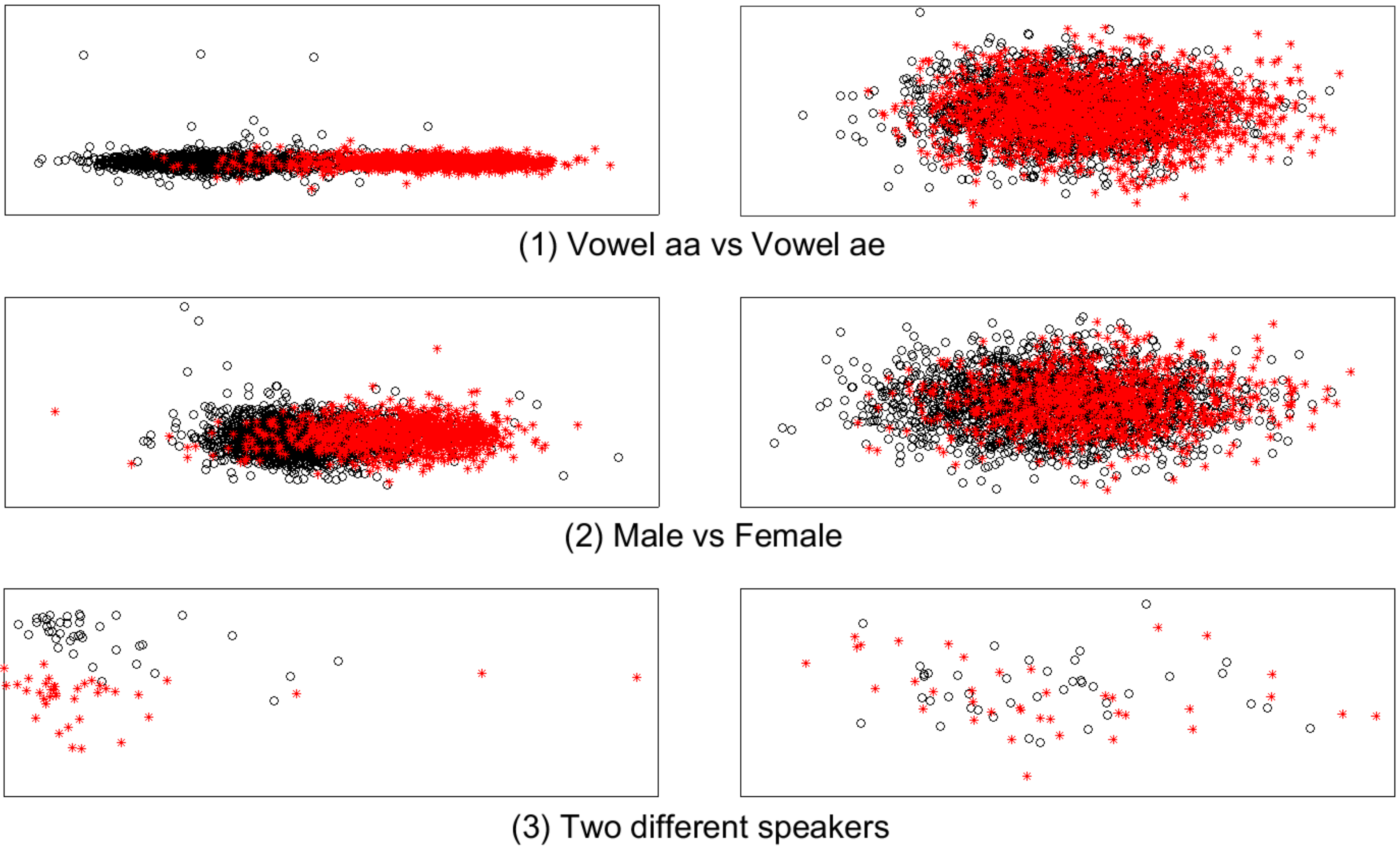}
\caption{Distribution with class labels in the observation space (left column) and the code space (right column). The two colors/shapes represent different vowels in the first row,
different genders in the second row, and different speakers in the third row.}
\label{fig:class}
\end{figure}

The last experiment examines if the discriminative information of the observations is preserved in the code space.
To gain this purpose, we train an LDA to select the two most discriminative directions (for a particular classification task)
and plot the samples in both the observation space and code space. Three classification tasks are
investigated: (1) vowel $aa$ vs $ae$; (2) male vs female; (3) two different speakers. The results are shown in Fig.~\ref{fig:class}.
It can be seen that the class information is largely lost when transforming to the code space. This unwanted
property for most speech processing tasks, however, seems not surprising, as
the flow model tries to compress all the codes into a Gaussian ball which is compact and dense, without any class information
taken into account.

\section{Conclusions}
\label{sec:Cons}

We presented a preliminary study on the properties of the latent space derived by a normalization flow model for speech segments. The experimental results showed that
this code space possesses a favorable pseudo linear property, which means that perceptually important factors such as phonetic content and speaker traits can be changed gradually
by moving in the code space following a particular direction. This provides an interesting way of unsupervised speech factorization, where each salient factor corresponds to a particular
direction in the code space. Potential applications of this factorization include voice conversion
and noise cancellation. Future work will conduct more thorough studies on large databases and continuous speech. Another work
will investigate discriminative flow models which take class information into consideration.

%\section{REFERENCES}
%\label{sec:refs}

% References should be produced using the bibtex program from suitable
% BiBTeX files (here: strings, refs, manuals). The IEEEbib.bst bibliography
% style file from IEEE produces unsorted bibliography list.
% -------------------------------------------------------------------------
\newpage
\bibliographystyle{IEEEbib}
\bibliography{refs}

\end{document}